\documentstyle[epsfig]{jaa}
\begin{document}

\title[High Mass X-ray Binary Pulsar 4U~0114+65]
{Luminosity dependent study of the High Mass X-ray Binary Pulsar 4U~0114+65 with 
\it ASCA}

\author{U. Mukherjee  and B. Paul}

\date{}

\maketitle

\begin{center}
Tata Institute of Fundamental Research,\\
Homi Bhabha Road, Colaba, Mumbai - 400 005\\~\\

\end{center}

\begin{abstract}

 Here we report the spectral characteristics of the high and low states
 of the pulsar 4U~0114+65 and examine the change in the parameters of the spectral 
 model. A power law and a photoelectric absorption by material along the line of 
 sight together with a high energy cut-off suffice to describe 
 the continuum spectrum in both the states. 
 A fluorescence iron line at $\sim$6.4 keV 
 is present in the high as well as in the low state, though it is less intense
 in the latter. The photon index, cut-off energy and e-folding energy values hardly 
 show any discernible change over the states. 
 We compare these 
 spectral characteristics as observed with ASCA to that with other satellites.
 We also compare the spectral characteristics of 4U~0114+650 with other X-ray sources
 which show intensity variation at different time scales.
 
\end{abstract}

{\bf keywords} stars: pulsar: individual (4U~0114+65) --- X-rays:pulsar

\section{Introduction}

4U~0114+650 is a pulsar which belongs to the class of High Mass X-ray Binaries (HMXB). 
It was discovered in the SAS-3 Galactic Survey (Dower \& Kelley 1977).
The pulsar has a B1Ia supergiant optical companion and the binary system is at a distance 
of $\sim$7 kpc (Reig et al. 1996). Crampton, Hutchings, \& Cowley (1985) reported an
orbital period of $\sim$11.59 days for this HMXB from the optical radial velocity measurements.
Their measurements were not able to distinguish between a circular or elliptical orbit for
this binary. Corbet et al. (1999) obtained X-ray intensity modulations at a period of 
$\sim$11.63 days from the long term light curve of RXTE-ASM.
In addition to the orbital modulation, the X-ray light curve of this source shows 
considerable variability, with flaring
activity for a few hours (Apparao et al. 1991, Finley et al. 1992) and also short-term
flickering for minutes (Koenigsberger et al. 1983). The presence of a $\sim$2.8 hr
periodicity has been observed in the X-ray light curves of 4U~0114+65 from the analysis
of archival EXOSAT and ROSAT data (Finley et al. 1992).
This has been re-confirmed by Corbet et al. (1999) with the data from RXTE-ASM and Hall et al. (2000)
with RXTE-PCA. This periodicity has been interpreted as the spin period of the pulsar.
Very recently, Farrell et al. (2005) reported the detection of a superorbital
period of $\sim$30.7 days in this pulsar by analyzing the data of RXTE-ASM.

The X-ray spectrum emanating from the 
pulsar is nicely fitted with a generic model applicable in the case of HMXBs
consisting of a power law with an absorption along the line of sight,
modified at high energies by an exponential cut-off. There is also the presence of 
a fluorescent line at $\sim$6.4 keV due to neutral or lowly ionized iron 
(Yamauchi et al. 1990, Masetti et al. 2005).
The typical values of the model parameters are a photon index $\sim$1, the cut-off
energy at $\sim$8 keV and the e-folding energy at $\sim$20 keV. The column density 
values vary from $\sim$3 $\times$10$^{22}$ atoms cm$^{-2}$ (Hall et al. 2000, with 
RXTE) to $\sim$15 $\times$10$^{22}$ atoms cm$^{-2}$ 
(Masetti et al. 2005, with BeppoSAX). X-ray spectroscopy of this source
during the different intensity states has been carried out in different energy bands 
with data from various satellites. Like, Koenigsberger et al. (1983) with 
combined data from Einstein Observatory, HEAO-1 and OSO-8, Yamauchi et al. (1990) with
GINGA, Apparao et al. (1991) with EXOSAT, Hall et al. (2000) with RXTE, Bonning \& Falanga (2005) 
with INTEGRAL and Masetti et al. (2005) with BeppoSAX.

A very preliminary overall X-ray spectral analysis of this source observed by ASCA
was reported earlier (Ebisawa 1997). The preliminary analysis showed the presence of 
a fluorescent iron-line at $\sim$6.4 keV and reported a variable 
hydrogen column density (N$_H$). But the broad-band spectral characteristics of 4U~0114+65
has now been established with BeppoSAX which confirms the presence of a high energy cut-off
in the spectra in both the states. Since the cut-off energy 
falls within the ASCA energy band, we have reanalyzed the ASCA data and determined the 
column density and iron-line parameters accurately.
Since ASCA has a better spectral resolution compared to other detectors which have 
observed this source and it also has a low energy coverage comparable to BeppoSAX,
it is imperative to investigate the source in detail with ASCA. 
In this paper, we map the spectral change of 4U~0114+650 with ASCA in two
different intensity levels.
We also compare the spectral characteristics of this HMXB with the results 
obtained from other satellites at different orbital phases. For this purpose, we have 
described the observations in section 2, the data analysis and results in section 3 and 
thereafter discuss our results in the last section.

\section{Observations}

The present observation of the pulsar 4U~0114+65 was carried out by 
ASCA on 1997-02-10 (at an orbital phase of 0.19) for a useful exposure
time of $\sim$25 ks. The time span between the start and the end of the
observation was $\sim$55 ks, which is about 5.5\% of the orbital period. ASCA has 
two gas-imaging spectrometers (GIS) and two solid-state imaging spectrometers 
(SIS). The energy resolutions are 500 eV
and 130 eV (FWHM) @ 6 keV for the GIS and SIS detectors respectively. 
For more details about ASCA, the reader is referred to Tanaka, Inoue, \& Holt (1994).
During this observation, the GIS was operated in the standard PH mode 
and the SIS was used in the Bright mode.

\section{Data Reduction, Analysis and Results}

For this work, we have used the standard data reduction 
procedure of the ASCA guest observer facility. We have taken the 
archival screened data for our analyses. The source light curves and the spectra were obtained 
from circular regions of radii of 6$'$ and 3$'$ around the centroid of the source
from the GIS and SIS respectively. The background light curves and spectra were 
extracted from the source-free part of the field-of-view from similar circular regions. 
For better statistics, the light curves and spectra from the GIS2 
and GIS3 detectors, as well as the data from the SIS0 and SIS1 detectors were 
combined together. A detailed description of ASCA data reduction and analysis can be found 
in, http://heasarc.gsfc.nasa.gov/docs/asca/abc/abc.html

\subsection{The Light Curve}

We have shown the total GIS and SIS background subtracted light curves with a 
binsize of $\sim$100 s in Figure 1. The light curve shows flaring state during 
the first 20 ks and low state afterwards.
On the average, there is a decay of the count rate from 10 ks onwards 
as can be seen from the light curves. 
The intensity at the peak of the flare is $\sim$12--15 times than the persistent 
low level emission. At times, the intensity almost drops down to zero (at $\sim$22 
and $\sim$32 ks in Figure 1). In addition, there are 
short time-scale intensity variations for a few minutes. For our 
analyses, we have divided the whole light curve into two sections. 
The first section correspond to the  high state and is between 0--18 ks. The other 
half is ascribed to the low state of the source. 
The GIS and SIS average count rates (background subtracted) for the whole observation are 3.9
and 3.0 count s$^{-1}$ respectively. The background subtracted count rates for the high state 
GIS and SIS light curves are 7.8 and 6.5 count s$^{-1}$ respectively, whereas the low state count rates for 
the said instruments are 2.0 and 1.7 count s$^{-1}$. We were not able to detect the $\sim$2.7 hr
periodicity from the light curves. Strong aperiodic variability at short and long time scales
and inadequate exposure prohibits us from finding pulsations at the reported 
2.7 hr. spin period. 

\subsection{The Spectral Model and the Fitting}

The spectral fitting was carried out in XSPEC version 11.2.0 (Shafer, Haberl $\&$ Arnaud 1989). 
After appropriate background subtraction, the GIS
and SIS spectra were fitted simultaneously.
The energy bands used for the fitting was 0.7--10.0 keV for the GIS and 0.5--9.0 keV
for the SIS, where the effective areas and energy responses of the detectors are well defined.
The 1024 channels of the GIS and the 512 channels of the SIS spectra were
grouped suitably. In all of our fits described below, we have included an interstellar
photoelectric absorption along the line of sight ({\it wabs} in XSPEC).
A constant factor was incorporated to account for the relative normalization 
of the GIS and SIS detectors. 

We extracted the spectra for the  high and the low intensity levels (as defined above) respectively.
The  high state spectrum was first fitted with a simple power law and residuals were 
seen between $\sim$6.2--6.6 keV. This 
prompted us to include a Gaussian line in the model. And that expectedly improved the fit  
from a $\chi^{2}$ of 417 for 265 degrees of freedom to a $\chi^{2}$ 
of 318 for 262 degrees of freedom.
Though the reduced $\chi^{2}$ of 1.2 was reasonably acceptable, there apparently remained some residuals 
between 1.0--3.0 keV. Also from $\sim$7 keV onwards there was a hint of a high energy cut-off 
in the residuals. Hence we added the high energy cut-off in our model.
That helped to weed out the above mentioned features in the residuals altogether.
Figure 2 shows the high state spectrum along with the residuals.
The low state spectra was fitted likewise  
and we obtained an acceptable fit without any feature in the residuals.
The low state spectrum is shown in Figure 3 with the best fit model and the 
residuals.

After fitting the high and low state spectra as above, we tried to 
fit the total spectrum (for the full time duration) with the same model.
Though we obtained a reasonable reduced $\chi^{2}$ of $\sim$1.2, the parameters,
except the line-centre energy, could not be constrained. 
and the best fit model shows structured residuals between 1.0--3.0 keV (Figure 4). 
In table 1, we give the best fit model parameters with the 90\% error estimates.
For the composite spectrum, the model parameters are not constrained well
and error estimates are not given.
Continuum parameters are identical within error-bars while the line flux is 
lower by a factor of $\sim$4.
The e-folding energy is not well constrained in the high state. This is possibly 
due to the limited energy range offered by ASCA. But in the low state, it is 
comparatively better constrained.

\section{Discussion}

The main objective of our work was to measure the spectral 
parameters of the pulsar 4U~0114+65 in its high and low states
with the now known model components that fit the broad band spectrum well
and to study the iron emission line parameters vis-a-vis the column density (N$_H$)
and continuum parameters at different orbital phases and intensity levels.
As enunciated in the results, the most suitable model 
describing the source in the different intensity states with ASCA was found to be
a power law modified by a cut-off, along with an exponential absorption along the 
line of sight. The presence of a Gaussian line at $\sim$6.4 keV, the fluorescent iron 
line, was also confirmed in both the spectra. 
It can be seen that as the unabsorbed luminosity between the two 
states varies by a factor of $\sim$4.5, there is a little change in the 
spectral parameters of the continuum, i.e. in the photon index 
and the column density ($N_{H}$). The most significant difference between the high
and low state spectra is a change in the iron-line flux by a factor similar 
to the change in the continuum. 

It is pertinent to compare our results with those obtained from other satellites.
In order to do that, we have projected in a nutshell in Table 2, the spectral parameters
of this pulsar as measured with other satellites. We have also mentioned the orbital
phases at which those observations were made to examine if there is any orbital phase
dependence of the spectral parameters. The orbital period and the epoch was taken from
Corbet et al.  (1999). We have specifically mentioned the high and low states in the
table to bring out any possible differences.
                                                                                                                            
The RXTE and GINGA observations found marginal evidence for an increase of $N_{H}$
during the low states while the BeppoSAX observations found $N_{H}$ to increase by
a factor of 2 in the low state. We note that in any of the three cases mentioned
above, the increase in $N_{H}$ by itself cannot explain occurrence of the low state.
The galactic column density along the line of sight of 4U~0114+65 is $\sim$0.8
$\times$10$^{22}$ cm$^{-2}$. This is about an order of magnitude lower than the column
density values of 4U~0114+65 obtained previously or in the present work.
The $N_{H}$ values measured with ASCA seem to compare well with that obtained with
RXTE and GINGA, though the ASCA observation shows a slight decrease in the value of
$N_{H}$ in the low state, and the $N_{H}$ values in the present work are a
factor of $\sim$2--4 times lesser than that measured with BeppoSAX.
From Table 2, if we compare only the high state
observations or only the low state observations, we do not find any clear orbital
phase dependence of the N$_{H}$ or the equivalent width of the iron line. If the
soft X-ray absorption is due to the material in the outflowing wind from the companion
star and if the iron emission line is produced by the reprocessing of the continuum
hard X-rays in this material, a smooth orbital phase dependence of the N$_{H}$ and
the equivalent width on orbital phase is expected. A similar finding in another 
pulsar GX~301--2 was interpreted in terms of a clumpy structure of the stellar wind 
(Mukherjee \& Paul 2004).

Regarding photon index of the power law component, Table 2 shows that apart from the
results from EXOSAT which point towards a very steep power law, the results from
other satellites generally indicate a $\Gamma$ of $\sim$1. The unusually high value
of $\Gamma$ with EXOSAT may be due to the slightly different spectral model used and
the limited band width. The values of $\Gamma$ obtained with ASCA is somewhat lower than 1. 
On the whole it is seen that $\Gamma$ for this pulsar does not change appreciably
with X-ray intensity.  

From Table 2, we notice that the equivalent width of the fluorescent
iron-line is comparatively higher in some of the low state observations
and on most occasions (with {\it BeppoSAX} and {\it RXTE}), the 6.4 keV line is
better detected in the low state. However with ASCA, we report the detection of
the iron-line in both the states with appreciable and similar equivalent widths (Table 2).
Since ASCA has the best spectral resolution in comparison to the other detectors
mentioned herein, the measurements of the iron line parameters is likely to be most
accurate with ASCA.
                                                                                                                            
A hint of a soft excess was reported in the BeppoSAX spectra of this source by
Masetti et al. (2005). They showed an apparent presence of the soft excess below 3 keV
which they fitted with a $\sim$ 0.3 keV thermal component.
However, in any of the ASCA fits covering an energy band lower compared to the reported
BeppoSAX spectrum, we did not find any signature of the so-called
soft excess as found in the spectra of many X-ray pulsars (Paul et al. 2002,
Hickox et al. 2004). Moreover, in the case of 4U~0114+65, for the unabsorbed flux 
observed with ASCA in the high state ($\sim$6$\times$10$^{-10}$ ergs cm$^{-2}$ s$^{-1}$) 
along with the N$_H$ of 3.4$\times$10$^{22}$ atoms cm$^{-2}$ falls in the region 
in Figure 1 of Hickox et al. (2004) which shows a collection of pulsars lacking 
soft excess.

During our observation, the transition of 4U~0114+650 between the high and the low state is abrupt. Similar
abrupt flux changes have been observed in another pulsar GX 1+4 which also shows aperiodic
transitions between its high and low states and thus requires special mention here.
Naik, Paul \& Callanan (2005) studied GX~1+4 for an extended period with observations by
BeppoSAX and reported an order of magnitude decrease in the X-ray flux for a part of the
observation, similar to the present observation of 4U~0114+650. In GX~1+4, the absorption
column density was an order of magnitude higher in the low state, which is not the case
for 4U~0114+650. In addition, Naik et al. (2005) found a strong increase in the iron line
equivalent width in the low state which is also different from what we have observed in
4U~0114+650.
                                                                                                                            
Equivalent width of the iron line is found to increase appreciably in the low
states of several other pulsars like Her X-1 (Naik \& Paul 2003),  
LMC~X-4 (Naik \& Paul 2004)and SMC X-1 (Vrtilek et al. 2005). The high and low states 
of the latter sources are part of
the super-orbital period arising due to a precessing warped inner accretion disk.
Though the pulsar 4U~0114+650 is
reported to have a super-orbital period (Farrell et al. 2005), the abrupt decrease
in X-ray flux over a few thousand second time scale reported here can not be assigned
to the precession of a warped accretion disk. The change in luminosity during our
observation can be ascribed to a local change in the density of the stellar wind from
which the pulsar accretes material. Rise and fall times of 1--2 ks as can be seen in
Figure 1, and an assumed orbital velocity of a few hundred km per sec indicates
clumpiness of the stellar wind at length scale of about 10$^{10-11}$ cm. The long term
light curve of this source measured with the RXTE-ASM does not have enough sensitivity
to measure incoherent intensity variations at the time scale of a few ks.
During the ASCA observation, the iron line flux
decreased with luminosity and has a direct correlation with the continuum luminosity
(Table 1). It supports a picture in which the amount of fluorescence material, like in the
form of a circumstellar shell at a large distance has remained unchanged but the local
accretion has changed.

\section{Conclusion} 

In this paper we have reported an accurate measurement of the spectral
parameters of the HMXB pulsars 4U~0114+650 in two different intensity levels. We have
detected iron emission lines of similar equivalent width in both the intensity levels
and the other spectral parameters are found to be similar in the two intensity levels.
This shows that the intensity variations at time scale of a few hours observed in
4U~0114+650 not due to increased absorption, but possiblly due to wind density that
causes changes in the accretion rate. We have compared the present observation of
this object with that done by other X-ray observatories and also with other HMXB pulsars
which show widely different intensity states. 4U~0114+650 is found to have characteristics
different from many other HMXB pulsars.

\section*{Acknowledgements}
This research has made use of data obtained from the High Energy Astrophysics
Science Archive Research Center (HEASARC), provided by NASA's Goddard Space
Flight Center. We would also like to thank the ASCA team for providing the 
data in the archive. UM would like to acknowledge the Kanwal Rekhi Scholarship 
of TIFR Endowment Fund for partial financial support.

\section*{References}

Apparao, K. M. V., Bisht, P., \& Singh, K. P. 1991, {\it ApJ}, {\bf 371}, 772 \\
Bonning, E. W., \& Falanga, M. 2005, {\it A\&A}, {\bf 436}, L31 \\  
Corbet, R. H. D., Finley, J. P., \& Peele, A. G. 1999, {\it ApJ}, {\bf 511}, 876 \\
Crampton, D., Hutchings, J. B., \& Cowley, A. P. 1985, {\it ApJ}, {\bf 299}, 839 \\
Dower, R., Kelley, R., Margon, B., \& Bradt, H. 1977, {\it IAUC}, {\bf 3144} \\
Ebisawa, K. 1997, {\it AAS}, 19111007 \\
Farrell, S. A., Sood, R. K., \& O'Neill, P. M. 2005, {\it astro-ph/0502008} \\
Finley, J. P., Belloni, T., \& Cassinelli, J. P. 1992, {\it A\&A}, {\bf 262}, L25 \\
Hall, T. A., Finley, J. P., Corbet, R. H. D., \& Thomas, R. C. 2000, {\it ApJ}, {\bf 536}, 450 \\
Hickox, R., C., Narayan, R., Kallman, T., R., 2004, {\it ApJ}, {\bf 614}, 881 \\
Koenigsberger, F., Swank, J. H., Szymkowiak, A. E., \& White, N. E. 1983, {\it ApJ}, {\bf 268}, 782 \\
Masetti, N., Orlandini, M., Dal Fiume, D., Del Sordo, S., Amati, L., Frontera, F., Palazzi, E., \&
Santangelo, A. 2005, {\it A\&A} (in press) {\bf astro-ph/0508451} \\
Mukherjee, U., \& Paul, B., 2004, {\it A\&A}, {\bf 427}, 567 \\
Naik, S., \& Paul, B., 2003, {\it A\&A}, {\bf 401}, 265 \\
Naik, S., \& Paul, B., 2004, {\it ApJ}, {\bf 600}, 351 \\
Naik, S., Paul, B., \& Callanan, P., J., 2005, {\it ApJ}, {\bf 618}, 866 \\
Paul, B., Nagase, F., Endo, T., et al., 2002, {\it ApJ}, {\bf 579}, 411 \\
Reig, P., Chakrabarty, D., Coe, M. J., Fabregat, J., Negueruela, I., Prince, T. A., Roche, P.,
\& Steele, I. A. 1996, {\it A\&A}, {\bf 311}, 879 \\
Shafer, R. A., Haberl, F., $\&$ Arnaud, K. A., 1989, {\bf XSPEC: An X-ray Spectral Fitting Package, ESA
TM-09 (Paris:ESA)} \\
Tanaka, Y., Inoue, H. \& Holt, S. S. 1994 {\it PASJ}, {\bf 46}, L37 \\
Vrtilek, S. D., Raymond, J. C., Boroson, B., \& McCray, R. 2005, {\it ApJ}, {\bf 626}, 307 \\
Yamauchi, S., Asaoka, I., Kawada, M., Koyama, K., \& Tawara, Y. 1990, {\it PASJ}, {\bf 42}, L53 \\
Yokogawa, J., Paul, B., Ozaki, M., et al., 2000, {\it ApJ}, {\bf 539}, 191 \\

\clearpage
\begin{figure}

\vskip 12. cm
\includegraphics{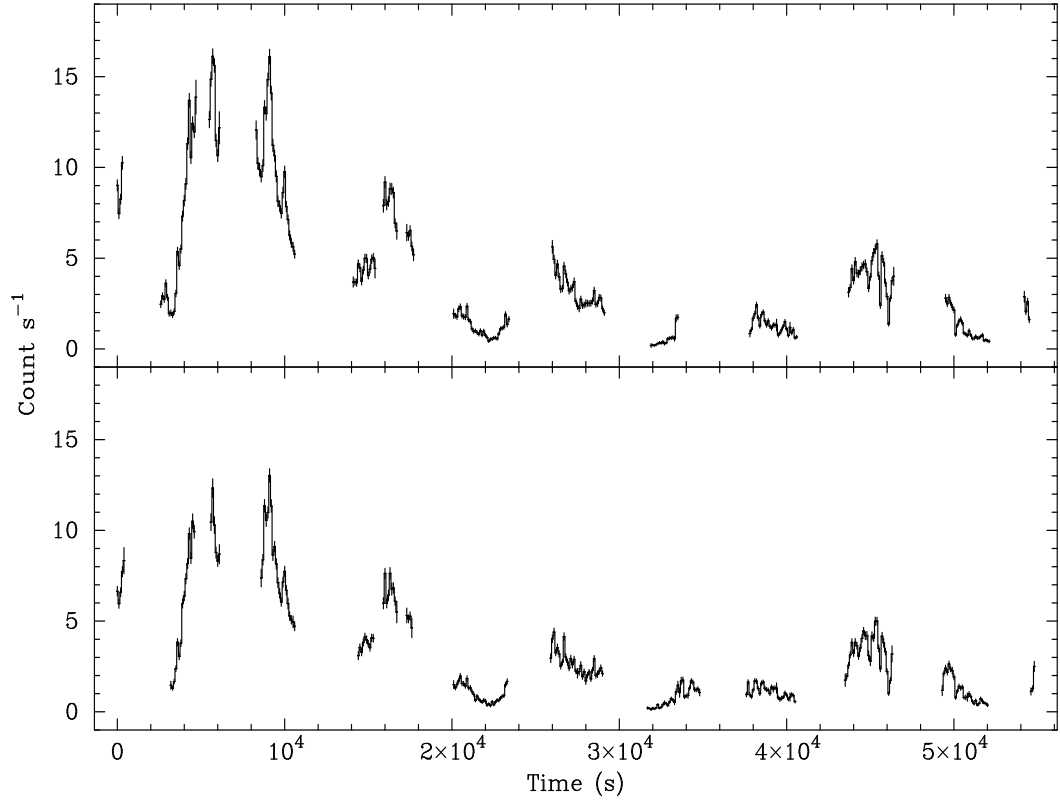}
\caption{The background subtracted light curves of 4U~0114+65 taken on
1997-02-10 with detectors onboard the ASCA observatory are shown here
with binsize of 100~s. The top panel shows the GIS light curve in the
0.7--10.0 keV band and the bottom panel shows the SIS light curve in
the 0.5--9.0 keV band.
}
\end{figure}
\clearpage

\begin{figure}
\vskip 12. cm
\includegraphics{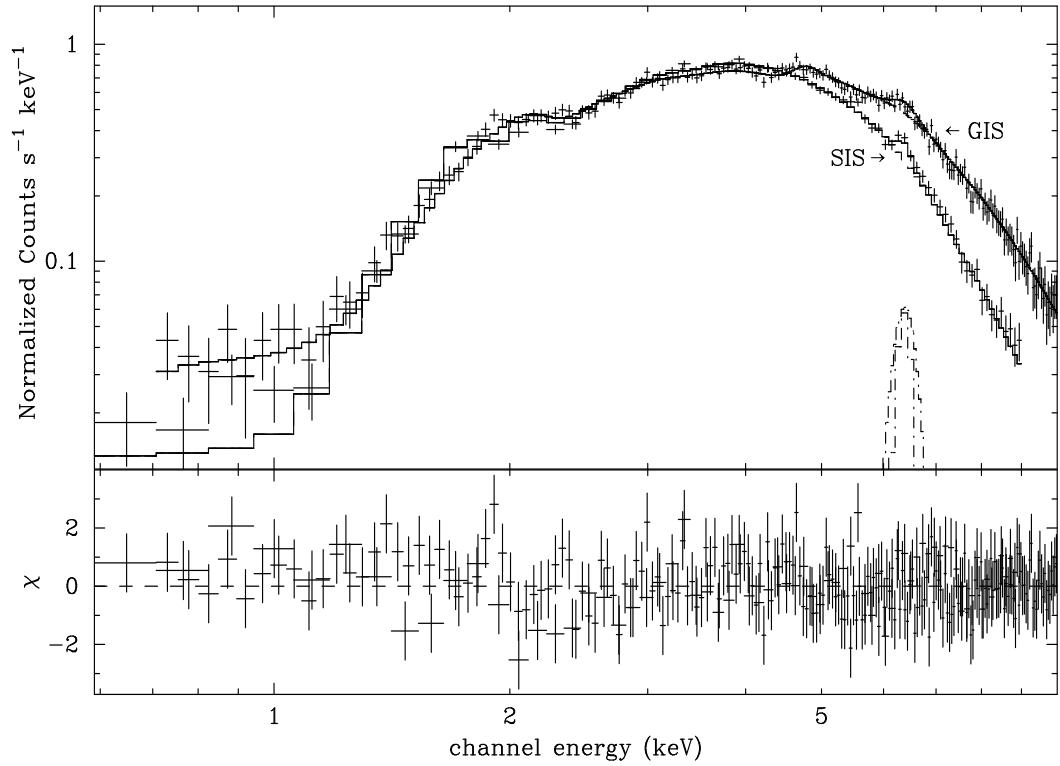}
\caption{The high intensity level X-ray spectrum of 4U~0114+65 measured
with the ASCA SIS and GIS are shown in the top panel along with the best
fitted spectral model components as histograms. Contribution of each data
point in the spectrum to the total $\chi^{2}$ is shown in the bottom
panel.}
\end{figure}
\clearpage

\begin{figure}
\vskip 12. cm
\includegraphics{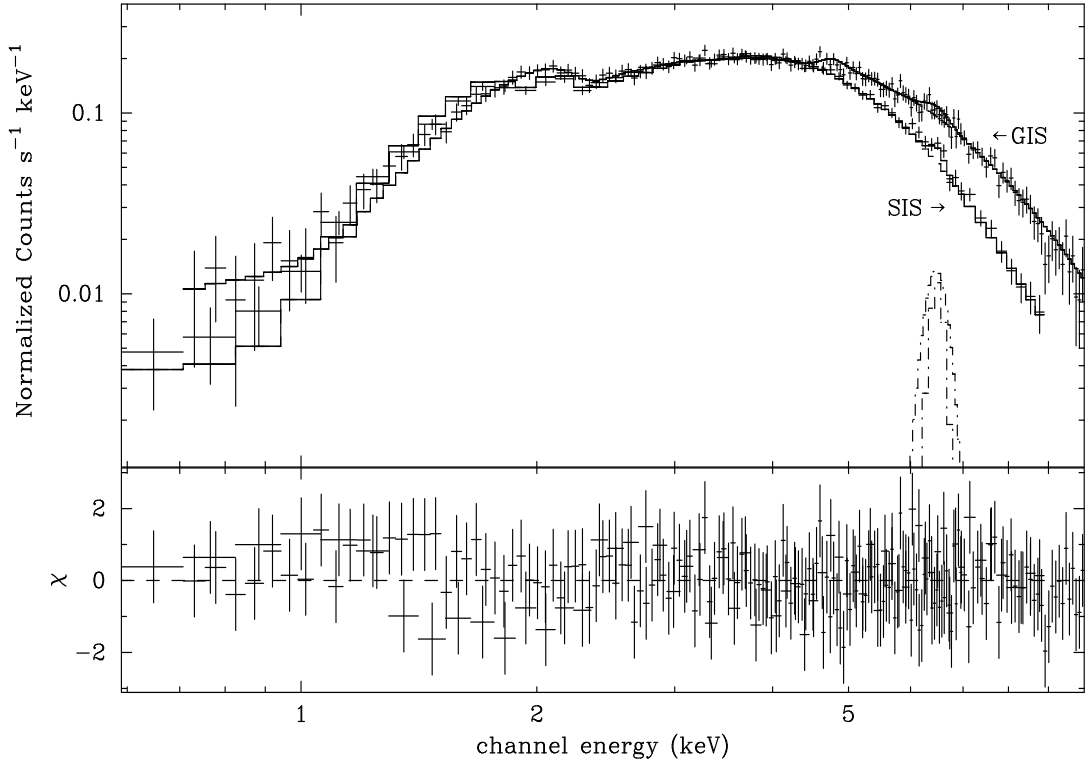}
\caption{Same as in Figure 2 for the spectrum at low intensity level.}
\end{figure}
\clearpage

\begin{figure}
\vskip 12. cm
\includegraphics{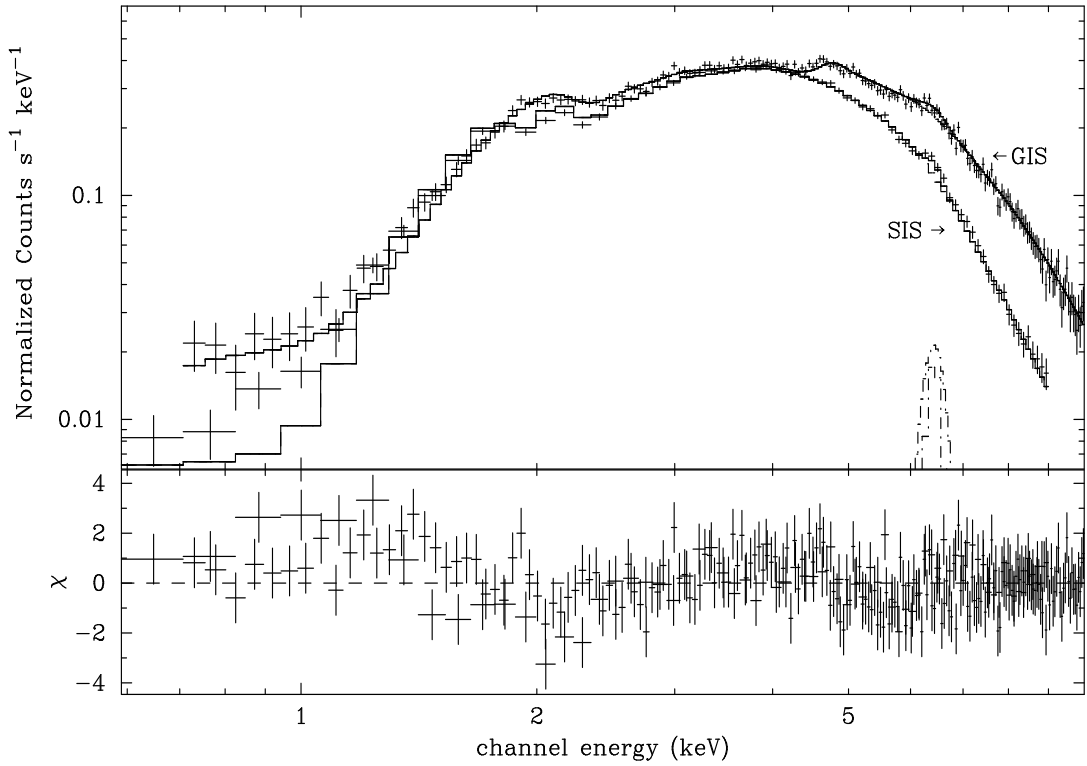}
\caption{The spectrum measured during the entire observation is shown here
with all other details same as in Figure 2. Unlike the spectra in the low
and high intensity levels (Figures 2 \& 3), the residuals show systematic
differences between the data and the best fitted spectral model.}
\end{figure}
\clearpage

\begin{table}
\caption{The Spectral Parameters of the States with 90$\%$ error bars (N$_H$ : Column Density, 
PL Norm : Power Law Normalization \& L$_{2-10~keV}$ : Unabsorbed Luminosity in the 2 - 10 keV range)}
\vskip 0.5cm
\begin{tabular}{ccccc}
\hline
\\
Parameters & High & Low & Total \\
\\
\hline
\\
N$_H$ (10$^{22}$ cm$^{-2}$) & 3.44$^{+0.13}_{-0.12}$ & 2.01$^{+0.14}_{-0.18}$ & 2.81 \\
\\
$\Gamma$ (Photon Index) & 0.64$^{+0.05}_{-0.05}$ & 0.49$^{+0.11}_{-0.15}$ & 0.62 \\
\\
PL Norm (photons keV$^{-1}$ cm$^{-2}$ s$^{-1}$) & 0.03$^{+0.002}_{-0.002}$ & 0.005$^{+0.001}_{-0.001}$ & 0.013 \\
\\
Line Energy (keV) & 6.38$^{+0.03}_{-0.06}$ & 6.49$^{+0.08}_{-0.13}$ & 6.43$^{+0.07}_{-0.06}$ \\ 
\\
Line Width (keV) & 0.006$^{+0.11}_{-0.006}$ & 0.013$^{+0.24}_{-0.003}$ & 0.016 \\
\\
Line Flux (10$^{-4}$ photons cm$^{-2}$ s$^{-1}$ ) & 5.60$^{+11.76}_{-1.55}$ & 1.63$^{+9.16}_{-0.70}$ & 2.07 \\
\\
Equivalent Width (eV) & 64.5 & 73.0 & 49.5 \\
\\
Cut-off Energy (keV) & 6.47$^{+0.38}_{-0.35}$ & 4.41$^{+0.37}_{-0.28}$ & 6.17 \\
\\
e-folding Energy (keV) & 8.54$^{+4.45}_{-1.65}$ & 7.82$^{+0.84}_{-1.04}$ & 7.83 \\
\\
Red. $\chi^{2}$ (d.o.f.) & 0.86 (260) & 0.65 (219) & 1.18 (260) \\
\\
L$_{2-10~keV}$ (10$^{36}$ ergs s$^{-1}$) &  3.65 & 0.80 & 1.71 \\
\\
\hline
\hline\\
\end{tabular}
\end{table}

\begin{table}
\caption{The Spectral Parameters obtained from various satellites. The High and Low 
states being designated as H and L respectively. 
(N$_H$ : Column Density, $\Gamma$ : Power Law Index, E$_c$ : Cut-off Energy, 
E$_f$ : e-folding Energy, EW : Equivalent Width of the 6.4 keV Fe line, L : Luminosity) 
\& 
($\phi$ : Orbital Phase, S : Satellite, A : {\it ASCA}, B : {\it BeppoSAX}, 
I : {\it Integral}, R : {\it RXTE}, EX : {\it EXOSAT}, G : {\it GINGA}, EN : {\it Einstein})
}
\vskip 0.5cm
\begin{tabular}{ccccccc}
\hline
\hline
\\
$\phi$ (S) & N$_H$  & $\Gamma$ & E$_c$ & E$_f$ & EW  & L  \\
\\
           & (10$^{22}$ cm$^{-2}$) & & (keV) & (keV) & (eV) & (10$^{36}$ ergs s$^{-1}$) \\
\\
\hline
\hline
\\
0.19 (A) & 3.44$^{+0.13}_{-0.12}$ & 0.64$^{+0.05}_{-0.05}$ & 6.47$^{+0.38}_{-0.35}$ & 8.54$^{+4.45}_{-1.65}$ & 64.5 & 3.65 (H) \\
\\
         & 2.01$^{+0.14}_{-0.18}$ & 0.49$^{+0.11}_{-0.15}$ & 4.41$^{+0.37}_{-0.28}$ & 7.82$^{+0.84}_{-1.04}$ & 73.0 & 0.80 (L)\\
\\
0.67 (B) &  9.7$^{+0.7}_{-0.9}$ & 1.33$^{+0.09}_{-0.16}$ & 12$^{+2.0}_{-3.0}$ & 21$^{+4.0}_{-3.0}$ & -- & 1.35 (H) \\
\\
         &  15.4$^{+2.0}_{-1.7}$ & 0.9$^{+0.2}_{-0.2}$ & 6.0$^{+0.9}_{-0.7}$ & 17$^{+5.0}_{-3.0}$ & 80 & 0.37 (L) \\
\\
0.85 (I) & -- & 1.6$^{+0.5}_{-0.5}$ & 9.0$^{+14.1}_{-8.8}$ & 22.1$^{+12.1}_{-6.0}$ & -- & 1.8 (H) \\
\\
0.82 (R) & 3.2$^{+0.3}_{-0.3}$ & 1.28$^{+0.04}_{-0.05}$ & 8.1$^{+0.4}_{-0.4}$ & 19.2$^{+1.1}_{-1.0}$ & & 1.25 (H) \\
\\
         & 4.1$^{+0.4}_{-0.4}$ & 1.63$^{+0.04}_{-0.02}$ & 11.3$^{+0.8}_{-0.9}$ & 22.4$^{+6.7}_{-4.7}$ & 250 & 0.2 (L) \\
\\
0.35 (EX) & 9.4$^{+1.9}_{-1.9}$ & 2.47$^{+0.30}_{-0.30}$ & -- & -- & 526 & 0.45 (L) \\
\\
0.36 (G) & 2.9$^{+0.2}_{-0.2}$ & 0.92$^{+0.02}_{-0.02}$ & 7.1$^{+0.3}_{-0.3}$ & 16.1$^{+0.8}_{-0.8}$ & 70 & 5.0 (H) \\
\\
         & 4.7$^{+0.6}_{-0.6}$ & 1.07$^{+0.03}_{-0.03}$ & 6.8$^{+1.0}_{-1.0}$ & 15.4$^{+4.0}_{-4.0}$ & 340 & 0.33 (L) \\
\\
0.78 (EN) & 3.2$^{+1.0}_{-1.0}$ & 1.2$^{+0.1}_{-0.1}$ & -- & -- & -- & 2.11 (H) \\
\\
          & 3.2$^{+1.0}_{-1.0}$ & 1.2$^{+0.3}_{-0.3}$ & -- & -- & -- & 0.12 (L) \\
\\ 
\hline
\hline \\
\end{tabular}

** The Luminosities quoted in both the Tables have been calculated with 
~~the distance to the pulsar as 7 kpc. **
\end{table}   

\end{document}